# Emergence of superconductivity in single-crystalline LaFeAsO under simultaneous Sm and P substitution

Nikolai D. Zhigadlo [a,*], Roman Puzniak [b], Philip J. W. Moll [c], Fabio Bernardini [d], Toni Shiroka [e,f,†]

[a] *CrystMat Company, CH-8037 Zurich, Switzerland*

[b] *Institute of Physics, Polish Academy of Sciences, Aleja Lotnikow 32/46, PL-02668 Warsaw, Poland*

[c] *Max Planck Institute for the Structure and Dynamics of Matter, 22761 Hamburg, Germany*

[d] *Dipartimento di Fisica, Università di Cagliari, IT-09042 Monserrato, Italy*

[e] *Paul Scherrer Institute, CH-5232 Villigen PSI, Switzerland*

[f] *Laboratory for Solid State Physics, ETH Zurich, CH-8093 Zurich, Switzerland*

## Abstract

We report on the high-pressure growth, structural characterization, and investigation of the electronic properties of single-crystalline LaFeAsO co-substituted by Sm and P, in both its normal- and superconducting states. Here, the appearance of superconductivity is attributed to the inner chemical pressure induced by the smaller-size isovalent substituents. X-ray structural refinements show that the partial substitution of La by Sm and As by P in the parent LaFeAsO compound leads to a contraction in both the conducting $Fe_2(As,P)_2$ layers and the interlayer spacing. The main parameters of the superconducting state, including the critical temperature, the lower- and upper critical fields, as well as the coherence length, the penetration depth, and their anisotropy, were determined from magnetometry measurements on a single-crystalline $La_{0.87}Sm_{0.13}FeAs_{0.91}P_{0.09}O$ sample. The critical current density ($j_c$), as resulting from loops of magnetization hysteresis in the self-generated magnetic field, is $2 \times 10^6$ A/cm$^2$ at 2 K. Overall, our findings illustrate a rare and interesting case of how superconductivity can be induced by co-substitution in the 1111 family. Such approach delineates new possibilities in the creation of superconductors by design, thus stimulating the exploration of related systems under multi-chemical pressure conditions.

Corresponding authors.

[*] nzhigadlo@gmail.com (N.D. Zhigadlo); http://crystmat.com

[†] tshiroka@phys.ethz.ch (T. Shiroka)

# 1. Introduction

The application of high pressure is currently one of the most promising routes towards the synthesis and control of novel materials, most prominently of high-temperature superconductors [1,2]. Unfortunately, not always the transient phenomena and materials emerging under applied pressure persist also when the external (*i.e.*, physical) pressure is removed. A more effective alternative for stabilizing new materials is the use of internal (*i.e.*, chemical) pressure, typically achieved through the introduction of new chemical species in a solid compound [3,4]. Although chemical pressure mimics closely the effects of physical pressure, its outcome is often very different [5]. The origin of this difference lies in the inhomogeneous character of chemical pressure which, besides inducing an overall compression of the structure, it expands selectively certain regions of the unit cell. In this respect, studies that elucidate how the crystal- and electronic structure of compounds evolve under chemical pressure are important in the search for new superconductors and in developing a deeper understanding of the mechanisms involved in the superconductivity.

In Fe-based compounds, superconductivity is known to occur in the proximity of a suppressed antiferromagnetic (AFM) phase and to be tuned by carrier doping, external-, and internal (chemical) pressure [6]. By applying pressure, one modifies the lattice parameters and, thus, changes the stability of the AFM long-range order and the density of carriers at the Fermi level. Currently, this strategy is widely adopted as an important way to discover new superconductors or to optimize the already known ones [7].

A distinctive feature of the Fe-based superconductors is the possibility to induce superconductivity by applying chemical pressure, even with a partial isovalent substitution of the smaller P ions for larger As ions. Since the parent materials of Fe-based superconductors are metals, they already contain a sufficient density of carriers to enable superconductivity, even without external charge doping. Hence, the primary effect of isovalent substitution is the weakening of the AFM state. Indeed, several theoretical [8,9] and experimental [10-12] studies have addressed this aspect and confirm such interpretation. The general consensus is that, while P substitution for As does not induce appreciable changes to the electron density, it influences significantly the localization of the hybridized states, the electronic bandwidth, and the topology of the Fermi surface. The key feature of the isovalent substitution is the preservation of the electron- and hole concentration of the parent compound also at other $x$ values ($As_{1-x}P_x$). Hence, isovalent substitution is expected to tune the magnetic character without changing the charge-carrier concentration. Yet, in reality, several 122 and 1111 systems show diverse responses upon P substitution. Thus, P substitution induces superconductivity in $BaFe_2As_2$ [13], $CaFe_2As_2$, $SrFe_2As_2$ [14], $EuFe_2As_2$ [15], and LaFeAsO [16], following the lowering of the tetragonal-to-orthorhombic structural transition temperature and the associated suppression of the spin-density-wave (SDW) transition. The $CeFeAs_{1-x}P_xO$ system, instead, exhibits a peculiar behavior: while initial studies found no evidence of superconductivity [17,18], latter reports showed superconductivity with $T_c \sim 4$ K within a narrow substitution region around $x \sim 0.3$ [19].

Within the homologous series of quaternary iron pnictides (1111-$Ln$Fe$Pn$O, $Ln$: lanthanide, $Pn$: pnictogen) the effect of internal chemical pressure on the $Ln_2O_2$ layers has also been studied. In LaFeAsO, the replacement of La with the smaller Y- or Sm ions produces chemical pressure and suppresses the SDW order, yet it does not induce superconductivity [20-22]. The simultaneous application of chemical pressure to both the $Fe_2Pn_2$ and $Ln_2O_2$ layers is rarely considered, not well understood, and it needs further theoretical- and experimental exploration. The main objective of our study was to evaluate the possibility of inducing superconductivity in the parent LaFeAsO compound by applying a dual-site chemical pressure, $i.e.$, to both the conducting- $Fe_2As_2$ and the charge-reservoir $La_2O_2$ layers, via a partial substitution of P ions for As and of Sm ions for La. The choice of this system was motivated by the extensive number of studies regarding single-site substitutions, which provide an ideal starting point for comparing our results with those of the widely investigated LaFeAs$_{1-x}$P$_x$O and La$_{1-x}$Sm$_x$FeAsO systems. In LaFeAs$_{1-x}$P$_x$O, phosphorus doping leads to the appearance of superconductivity in the $0.2 < x < 0.4$ region, with a maximum $T_c$ of 10.8 K at $x \approx 0.3$ [16]. Hence, for the present study, we selected a composition on the non-superconducting region of the LaFeAs$_{1-x}$P$_x$O phase diagram, $i.e.$, LaFeAs$_{0.9}$P$_{0.1}$O, and applied additional chemical pressure to the $La_2O_2$ layers by substituting 15 % of La ions with Sm. The application of this kind of dual-site chemical pressure led to mixed effects and to the appearance of superconductivity at 13.3 K in La$_{0.87}$Sm$_{0.13}$FeAs$_{0.91}$P$_{0.09}$O.

Note that, contrary to the hydrostatic pressure, which generally produces homogeneous effects, in complex compounds such as $Ln$1111-type oxypnictides, chemical pressure may selectively affect particular units. To date, despite extensive studies, the relation between the magnetic order, superconductivity, and crystallo-chemical parameters is still unclear. One of the causes for this is the lack of high-quality single crystals, crucial for identifying the microscopic structural changes. Therefore, an essential part of the present work was consisted in the growth of single-crystalline samples, which successively allowed us to detect even small changes in the lattice parameters in presence of chemical pressure. The basic parameters of the superconducting state and their anisotropies were then determined by means of magnetization- and transport measurements performed on La$_{0.87}$Sm$_{0.13}$FeAs$_{0.91}$P$_{0.09}$O single-crystalline samples. As shown in detail below, our findings provide important information for further theoretical- and experimental investigations regarding the appearance of superconductivity in Fe-based superconductors under simultaneous chemical substitution.

**2. Crystal growth and experimental details**

For the growth of (La,Sm)Fe(As,P)O single crystals, we used the cubic-anvil high-pressure and high-temperature technique. Earlier on, this method was successfully employed to grow superconducting $Ln$Fe$Pn$O oxypnictides ($Ln$: lanthanide, $Pn$: pnictogen) [23-26] and numerous other compounds [27-29]. The details of the experimental setup can be found in our previous publications [30,31]. The nominal composition of the precursor was $La_{0.85}Sm_{0.15}FeAs_{0.9}P_{0.1}O$. High-purity powders ($\geq$ 99.95%) of the starting materials LaAs, LaP, SmAs, $Fe_2O_3$, and Fe were weighed according to the stoichiometric ratio, thoroughly grounded using a mortar, and mixed with NaAs flux. For each growth attempt, typically we used ~ 0.45 g of $La_{0.85}Sm_{0.15}FeAs_{0.9}P_{0.1}O$ and ~ 0.2 g of NaAs. A pellet containing the precursor and the flux was enclosed in a boron-nitride container and placed inside a pyrophyllite cube with a graphite heater. Due to the toxicity of arsenic and the hygroscopic nature of NaAs, all procedures related to the sample preparation were performed in a glove box. An apparatus with six tungsten-carbide anvils compressed the cubic cell to 3.5 GPa at room temperature. The crystal growth was performed by heating up the mixture of solute and flux to the maximum temperature of ~ 1450 °C in 3 h. The mixture was kept there for 10 h, cooled first to 1250 °C at a rate of ~ 3.5 °C $h^{-1}$, held at this temperature for 10 h, and finally cooled down to room temperature in 3 h [Fig. 1(a)]. After completing the crystal growth, the crystalline products were immersed in distilled water to dissolve the remaining flux and then disaggregated by ultrasonic waves. As shown in Fig. 1(b), the crystals exhibit a platelet-like geometry, roughly 300-400 μm long in the *ab*-plane and ~ 20-30 μm thick along the *c*-axis. The stoichiometry of the as-grown crystals was determined by energy-dispersive x-ray spectroscopy (EDX, Hitachi S-3000 N), as resulting from the average of six measurements on each crystal. This was further confirmed by structural refinements, with both methods indicating that the actual composition of crystals is $La_{0.87}Sm_{0.13}FeAs_{0.91}P_{0.09}O$. A compositional spread of less than 3% demonstrates the good homogeneity of the Sm and P co-substituted single-crystalline samples.

The x-ray single-crystal diffraction measurements were performed at room temperature in an Oxford Diffraction SuperNova area-detector diffractometer [32] using mirror optics, monochromatic Mo $K_\alpha$ radiation ($\lambda$ = 0.71073 Å), and an Al filter [33]. The unit cell parameters were obtained from a least-squares refinement, using reflection angles in the range $2.53° < \theta < 51.11°$. The data were analyzed using the *CrysAlis$^{Pro}$* data collection and processing software [32]. The recorded intensities were corrected for the Lorentz- and polarization effects, while the absorption was corrected by using the SCALE3 ABSPACK multi-scan method in *CrysAlis$^{Pro}$* [32]. For the magnetization measurements, we employed roughly rectangular single-crystalline samples with approximate dimensions of 220 × 220 $μm^2$ in the *ab*-plane and about 23-μm thick along the *c*-axis. The crystal with the sharpest superconducting transition was then selected. The measurements were carried out in the 2 – 25 K temperature range, in magnetic fields up to 6.5 T, using a Magnetic Property Measurement System (MPMS-XL) by Quantum Design, equipped with a reciprocal sample option. The magnetic field was applied parallel to the crystal *c*-axis, as well as in the *ab*-plane (*i.e.*, perpendicular to the *c*-axis). Four-point resistivity measurements were performed in a 14-Tesla Quantum Design Physical Property Measurement System (PPMS). Micrometer-sized platinum (Pt) leads were precisely deposited onto a plate-like crystal using a focused ion beam (FIB) method without altering the bulk superconducting properties [34].

The electronic-structure calculations were performed using the all-electron code WIEN2k [35] based on the full-potential augmented plane-wave plus local orbitals method (APW+LO). Given the low concentration of the substitutional Sm and P dopants, the La$_{0.87}$Sm$_{0.13}$FeAs$_{0.91}$P$_{0.09}$O was simulated by using its LaFeAsO parent compound. For our calculations we employed the structural parameters determined experimentally (see Table I) and the Perdew-Burke-Ernzerhof (PBE) form of the generalized gradient approximation (GGA) [36]. We used muffin-tin radii of 2.50, 2.25, 2.10, and 1.90 a.u. for the La, Fe, As, and O atoms, respectively, and a plane-wave cut off $R_{MT} \times K_{max} = 7.0$. Finally, the integration over the Brillouin zone was done using a Monkhorst-Pack mesh of $16 \times 16 \times 7$ $k$-points for the self-consistent calculations, while a denser $64 \times 64 \times 32$ $k$-mesh was used to compute the Fermi surface within the conventional (primitive) cell.

## 3. Results and discussion

### 3.1. Structure modifications due to the simultaneous P-for-As and Sm-for-La substitution

In view of the often-antagonistic relationship between the magnetic order and superconductivity, a key question is: can one simultaneously suppress magnetism and induce superconductivity via double-site isovalent substitution? In fact, the magnetic order is known to be already suppressed by single-site P substitution [16]. Hence, one expects the emergence of superconductivity, e.g., in LaFeAs$_{0.9}$P$_{0.1}$O, by applying additional chemical pressure, e.g., via a further Sm-for-La substitution. This idea was triggered by our earlier observation of pressure-induced superconductivity in LaFeAsO$_{0.945}$F$_{0.055}$ [37], but now we replace external pressure with chemical pressure. Since this specific composition is at the boundary of the AFM and SC region, the application of even a moderate hydrostatic pressure of ~ 2.4 GPa leads to a strong increase in $T_c$ (from ~ 7 K to ~ 16 K) and diamagnetic response (from ~ 1% to ~ 35%). This fact strongly suggests that an optimal crystal structure is crucial to the emergence of bulk superconductivity.

To examine the effects of P- and Sm substitutions we carried out detailed single-crystal x-ray refinements on La$_{0.87}$Sm$_{0.13}$FeAs$_{0.91}$P$_{0.09}$O and compared our results with published data for the parent LaFeAsO compound, as well as with those of the single-site substituted LaFeAs$_{1-x}$P$_x$O and La$_{1-x}$Sm$_x$FeAsO. The lattice constants, atomic positions, and selected bond lengths and angles obtained from structural refinement of a single La$_{0.87}$Sm$_{0.13}$FeAs$_{0.91}$P$_{0.09}$O crystal, are summarized in Tables I and II. The refinement data demonstrate the good structural quality of the grown crystals. Then, we established quantitatively the amount of Sm and P atoms introduced into the structure. The resulting La/Sm and As/P atomic occupancies obtained from x-ray refinements were in a good agreement (within ~ 3 %) with the EDX analysis.

For an easy comparison of the substituted superconductors with the parent LaFeAsO compound, the key structural changes due to P- and Sm co-substitution are shown schematically in Fig. 2. The studied crystals adopt a ZrCuSiAs-type structure, consisting of stacks of two-dimensional Fe(As,P) and (La,Sm)O layers. While the interlayer bonds are of ionic character, the intralayer bonds are mostly of covalent nature. Here, $La^{3+}/Sm^{3+}$ is bonded in a four-coordinate geometry to four equivalent $O^{2-}$ and four equivalent $As^{3-}/P^{3-}$ atoms. $Fe^{2+}$ is bonded to four equivalent $As^{3-}/P^{3-}$ atoms to form a mixture of corner and edge-sharing Fe(As,P)$_4$ tetrahedra. $As^{3-}/P^{3-}$ is bonded in a eight-coordinate geometry to four equivalent $La^{3+}/Sm^{3+}$ and four equivalent $Fe^{2+}$ atoms. $O^{2-}$ is bonded to four equivalent $La^{3+}/Sm^{3+}$ atoms to form a mixture of corner and edge-sharing O(La,Sm)$_4$ tetrahedra. Such local environment makes it evident that significant changes in bond length, angle, and layer thickness will occur upon Sm and P substitution (see Table I and Fig. 2). Indeed, since the ionic radii of $Sm^{3+}$ (1.08 Å) and $P^{3-}$ (1.10 Å) are each smaller than those of $La^{3+}$ (1.16 Å) and $As^{3-}$ (1.19 Å) [38], it is to be expected that, upon Sm and P substitution, all the LaFeAsO lattice constants and, consequently, its unit-cell volume will decrease. Yet, we recall that the overall lattice-metrics response to the above co-substitutions is anisotropic. Compared to the unsubstituted LaFeAsO [$a$ = 4.0357(3) Å; $c$ = 8.7378(6) Å] [39], we find that, in a P- and Sm co-substituted sample, the lattice parameters change to $a$ = 4.0179(2) Å and $c$ = 8.6630(7) Å, respectively. This implies a $\Delta a/a \approx$ -0.44% and $\Delta c/c \approx$ -0.86%, in turn resulting in a unit-cell volume reduction by 1.73% compared to the pristine LaFeAsO.

The contraction along the $c$-axis is mainly due to the reduction of the As/P-Fe-As/P layer thickness [$S_2$ in Fig. 2], *i.e.*, to the reduction of the FeAs/P height, when As atoms are replaced by the smaller P atoms, whereas the La/SmO layer thickness remains essentially unchanged upon Sm substitution. These trends are very different from those observed in the single-substituted LaFeAs$_{1-x}$P$_x$O [16] and La$_{1-x}$Sm$_x$FeAsO systems [22]. Remarkably, in LaFeAs$_{1-x}$P$_x$O, the LaO layer thickness expands upon P substitution but, in La$_{1-x}$Sm$_x$FeAsO, it contracts upon Sm substitution, both outcomes clearly differing from the almost unchanged LaO layers in (La,Sm)Fe(As,P)O. These fundamental differences reflect the fact that the Sm-for-La substitution in (La,Sm)Fe(As,P)O compensates the expansion in the La-O layer caused by the P-for-As substitution, thus leaving the (La,Sm)O layer thickness almost unaffected. Interestingly, we find that, in La$_{0.87}$Sm$_{0.13}$FeAs$_{0.91}$P$_{0.09}$O, the spacing between the active Fe(As,P) and the spacer (La,Sm)O layers [$S_3$ in Fig. 2] decreases by 1.32% with Sm substitution, thus significantly strengthening the interlayer Coulomb interaction. This is indirectly confirmed by the stronger interlayer interactions in pristine SmFeAsO compared to LaFeAsO [22]. Our data are in good agreement with x-ray absorption investigations of La$_{1-x}$Sm$_x$FeAsO, where the spacing between the FeAs and LnO layers is shown to decrease with increasing Sm content, consistent with the thinner LnO layers in SmFeAsO [22].

Upon P substitution, the original Fe-As distance (2.4049 Å) in LaFeAsO decreases slightly (2.3972 Å). It is known that a reduced Fe-As/P distance may modify the hybridization between the Fe $3d$ and As $4p$ orbitals and, thus, quench the ordered Fe magnetic moments [40]. Theories based on antiferromagnetic spin fluctuations suggest that $h_{Pn}$, the pnictogen height above the iron plane, has a strong influence on the nature of the superconducting-order parameter [41]. Generally, pnictogen heights below the ~ 1.33 Å threshold value imply a nodal SC state [42]. Following these predictions, the $h_{Pn}$ = 1.308 Å of our crystals suggests a nodal-pairing scenario. However, one needs to bear in mind that disorder in $La_{0.87}Sm_{0.13}FeAs_{0.91}P_{0.09}O$ can arise from various sources, including substitution of different elements, site disorder, and structural defects, and can have a significant effect on the electronic properties of the material [43]. This conjecture warrants further experimental confirmation, ideally by techniques such as specific-heat measurements and angle-resolved photoemission spectroscopy. Further details on the evolution of the electronic-band structure for P- and Sm co-substituted LaFeAsO are reported in sec. 3.6.

To summarize, we studied the effects of chemical pressure in the spacer- and conducting layers of the LaFeAsO system upon the partial substitution of La with Sm and As with P. We find that, in the conducting FeAs layers, chemical pressure decreases the Fe-As distance and the Fe(As,P) layer thickness, whereas, in the spacer LaO layers, it drives the interlayer interactions, possibly increasing the overlap of the Fe $3d$ and As $4p$ orbitals. Such enhanced orbital overlap could account for the enhanced metallic character and, ultimately, for the onset of superconductivity in (La,Sm)Fe(As,P)O.

*3.2. Determination of the critical temperature*

Figure 3 presents the temperature dependence of the dc magnetization measured at 0.3 and 1 mT in the 2-25 K temperature range in zero-field cooling (ZFC) mode, with the magnetic field applied parallel to the $c$-axis. The critical temperature, here defined as the point where the $M_{ZFC}(T)$ curve deviates from a constant temperature-independent value, was found to be 13.3 K. The sharp transition to the superconducting state, with an almost linear dependence of the ZFC magnetic susceptibility below $T_c$, evidences the high quality of the studied crystal. The recorded magnetic response above $T_c$ is nearly zero, indicating that the crystal does not contain magnetic impurities, while the strong signal below $T_c$ is consistent with bulk superconductivity. A magnetic susceptibility significantly exceeding -1, as illustrated in Fig. 3, is due to the large demagnetizing field in the $H//c$-axis geometry. We note that the measured $T_c$ values in crystals from the same batch differ by less than 0.5 K, suggestive of a good microscopic homogeneity.

*3.3. Thermodynamic parameters - upper and lower critical fields*

We also examined the temperature dependence of the upper and lower critical fields in two geometries, *i.e.*, for *H*//*c*-axis and *H*//*ab*-plane. The temperature dependence of magnetization for selected fields, recorded in the *H*//*ab* geometry, is shown in Fig. 4(a). These data allowed us to determine $T_{c2}(H = \text{const})$, as the point where $M(T)$ deviates from linearity, here corresponding to the almost constant magnetic susceptibility in the normal state. The $T_{c2}(H)$ data obtained at various fields allowed us to plot the $H_{c2}(T)$ dependence for *H*//*c* and *H*//*ab*, as shown in Fig. 4(b). Here, in the vicinity of $T_c$, we note a minor change in the positive curvature of $H_{c2}(T)$. At higher fields (and lower temperatures), instead, $H_{c2}(T)$ changes linearly. For *H*//*c*, in the 1-3.5 T field range, we find $-\mu_0 dH_{c2}/dT = 0.76(8)$ T K$^{-1}$, whereas for *H*//*ab*, in the 1-6.5 T field range, we observe a steeper slope, 1.2(2) T K$^{-1}$. The anisotropy of the $-dH_{c2}/dT$ slope is about 1.6, in good agreement with an anisotropy of 2, determined from the ratio $H_{c2}^{//ab}/H_{c2}^{//c}$ in the 7-12 K temperature range [see inset in Fig. 4(b)]. Compared to the nearly optimally doped *Ln*1111 crystals, this anisotropy is very low near $T_c$, but it is quite close to that observed in underdoped PrFeAs(O,F) crystals ($T_c$ = 25 K) grown from the same flux [30]. In either case, the anisotropy is comparable to that of MgB$_2$ [44] and much smaller than that observed in the cuprates [45].

The zero-temperature value $H_{c2}(0)$ is normally obtained by extrapolating the $H_{c2}(T)$ data down to low temperatures. Here, the limited temperature range and the curvature of $H_{c2}(T)$ in the vicinity of $T_c$ (the latter reflecting the possible multiband nature of superconductivity), make such extrapolation difficult. Nevertheless, by assuming that $H_{c2}(0)$ is proportional to $T_c$ and to $-dH_{c2}/dT$, determined over a relatively wide field range (above the nonlinear region in the vicinity of $T_c$ [46]), we can estimate both $\mu_0 H_{c2}^{//c}$ (about 7 T) and $\mu_0 H_{c2}^{//ab}$ (about 11 T). These values correspond to a zero-temperature coherence length $\xi_{ab}$ = 6.9(2) nm and $\xi_c$ = 4.4(1) nm, respectively, as resulting from the relations [47]:

$$H_{c2}^{//c} = \Phi_0/2\pi \xi_{ab}^2, \tag{1}$$

$$H_{c2}^{//ab} = \Phi_0/2\pi \xi_{ab}\xi_c, \tag{2}$$

where $\Phi_0 = 2.07 \times 10^3$ T nm$^2$ is the magnetic flux quantum and $\xi_{ab}$ and $\xi_c$ are the coherence lengths in the *ab*-plane and perpendicular to it, respectively.

The temperature dependence of the lower critical field $H_{c1}$ was determined by identifying the field $H_{c1}^*$ at which the first vortices start to penetrate the sample at its surface, a value directly related to $H_{c1}$ [48]. As shown in Fig. 5(a), also here we measured the sample magnetization at different temperatures for magnetic fields applied parallel to the *ab*-plane and parallel to the *c*-axis.

For a given shape of the investigated crystal, the demagnetizing factor *D* was calculated for both the *H*//*ab* and *H*//*c* cases. The field $H_{c1}^*$ was estimated according to the procedure introduced in Ref. [48] and discussed in Ref. [49]. The quantity $(BV)^{1/2}$ was calculated from the measured magnetic moment $m = MV$ and plotted as a function of the internal magnetic field $H_{int} = H_{ext} - DM$. Here, $H_{ext}$ denotes the external magnetic field, $B$ the magnetic induction, and $V$ the sample volume. Since, in the Meissner state, $B = \mu_0(M + H_{int}) = 0$, from the $M(H_{int})$ data we can determine the field $H_{c1}^*$ above which the Meissner-state equation ceases to be valid. Considering that above $H_{c1}^*$ the magnetic induction $B$ empirically scales as the square of $H$, a plot of $(BV)^{1/2}$ as a function of $H_{int}$ allows a straightforward determination of $H_{c1}^*$. In case of a weak bulk pinning, the surface barrier may also play a role, since it determines the first field of flux penetration and the irreversibility line [50-52]. The surface barrier may also lead to asymmetric $M(H)$ loops, which make the descending branch almost horizontal. In our case, however, we observe symmetric magnetization loops. This means that only the bulk pinning controls the entry and exit of the magnetic-flux lines and, therefore, we can assume that $H_{c1}^* = H_{c1}$. The temperature dependence of $H_{c1}$, for both the $H//c$ and $H//ab$ directions, is shown in Fig. 5(b).

By extrapolating these data to zero temperature, we obtain the lower critical fields 0.85(9) mT for $\mu_0 H_{c1}^{//ab}$ and 13(2) mT for $\mu_0 H_{c1}^{//c}$ (see also Table 3). From the measured $H_{c1}(0)$ values, we can extract the magnetic penetration depth by means of the relations:

$$H_{c1}^{//c} = \Phi_0/4\pi\lambda_{ab}^2 \, [\ln(\kappa^{//c}) + 0.5], \tag{3}$$

$$H_{c1}^{//ab} = \Phi_0/4\pi\lambda_{ab}\lambda_c \, [\ln(\kappa^{//ab}) + 0.5]. \tag{4}$$

Here, $\lambda_{ab}$ and $\lambda_c$ denote the magnetic penetration depths related to the superconducting currents flowing in the $ab$-plane and along the $c$-axis, respectively, $\xi_{ab}$ and $\xi_c$ are the corresponding coherence lengths, and $\kappa^{//c} = \lambda_{ab}/\xi_{ab}$ and $\kappa^{//ab} = (\lambda_{ab}\lambda_c/\xi_{ab}\xi_c)^{1/2}$ are the Ginzburg-Landau parameters. For the iterative calculation also the zero-temperature $\xi_{ab}(0)$ and $\xi_c(0)$ values (here, equal to 6.9 nm and 4.4 nm, for the in-plane and out-of-plane coherence lengths), were used. The resulting magnetic penetration depths are: $\lambda_{ab}(0) \approx 225(30)$ nm and $\lambda_c(0) \approx 5000(700)$ nm. Table 3 provides a summary of all the superconducting-state parameters.

### 3.4. The critical current density

To evaluate the critical current density, we recorded the single-crystal hysteresis loops in magnetic fields applied along the $H//c$ and $H//ab$ directions at selected temperatures. The magnetization data for $H//c$ are shown in Fig. 6(a). By using the Bean's model [53,54] for a rectangular sample, we can estimate the superconducting critical current density from the formula:

$$j_c(H) = 2\Delta M(H)/a(1 - a/3b) \tag{5}$$

Here, $\Delta M$ (in A/m) is the width of the hysteresis loop, $a$ and $b$ are the sample dimensions in the plane perpendicular to the applied magnetic field, while $j_c$ is the critical current density. The resulting critical current densities vs. magnetic field, at different temperatures, for the magnetic-field geometries $H//ab$-plane and $H//c$-axis and are shown in Figs. 6(b) and 6(c), respectively. We note that the estimated critical current densities, $j_c$, are slightly lower than those of single-crystalline iron-pnictide superconductors with $T_c \sim 40$ K [55,56]. This is not surprising, considering that, the upper- and the lower critical fields of our crystals are smaller than those of the iron-pnictide counterparts and that the pinning is proportional to the thermodynamic critical fields.

### *3.5. Magneto-transport properties*

Figure 7 shows the temperature dependence of the magnetoresistance, from 2 to 300 K, for a single $La_{0.87}Sm_{0.13}FeAs_{0.91}P_{0.09}O$ crystal. Upon cooling in zero field, from room temperature to $T_c$ (onset – 15 K; 50% – 14 K; zero – 13 K), the resistance decreases by a factor of 2, thus defining the so-called resistivity ratio of the current sample. In the parent LaFeAsO compound, it is known that a drop in resistivity around 135 K is due to a structural transition associated with a spin-density-wave (SDW) type of antiferromagnetic order [57]. As shown in Fig. 7(b), also in our case we observe a weak (barely visible) change of slope near 135 K, which most likely is related to the spin-density-wave fluctuations.

Magnetoresistance measurements $\rho(T, H)$ near $T_c$ for magnetic fields parallel ($H//ab$) and perpendicular ($H//c$) to the FeAs/P-planes show a remarkably different behavior from that of SmFeAsO substituted with F for O, Co for Fe, or Th for Sm [24,25]. In all these cases, the magnetic field causes a significant broadening of the superconducting transition, but only a small shift of its onset, thus indicating a weak pinning and, accordingly, a large flux-flow dissipation. Conversely, in Sm- and P substituted La1111 crystals, the magnetic field shifts the onset of superconductivity to lower temperatures, but it essentially does not broaden the transition. The upper critical fields $H_{c2}//ab$ and $H_{c2}//c$, here defined as the fields where the resistivity is suppressed to 50% of $\rho_n$ (the zero-temperature extrapolation of the normal-state resistivity value), are shown in Fig. 8. The upper critical fields $H_{c2}$ of $La_{0.87}Sm_{0.13}FeAs_{0.91}P_{0.09}O$ increase linearly with decreasing temperature, with a slope $-\mu_0 dH_{c2}/dT = 6.4$ T K$^{-1}$ ($H//ab$) and $-\mu_0 dH_{c2}/dT = 1.96$ T K$^{-1}$ ($H//c$). These high slope values imply very high $H_{c2}(0)$ values.

As for the anisotropy, the La1111 structure is more anisotropic than that of 122 compounds, here manifested in a higher anisotropy of the upper critical field, $\gamma = H_{c2}^{//ab}/H_{c2}^{//c}$ [24-26]. The data in Fig. 8 suggest that, at temperatures sufficiently below $T_c$, the anisotropy $\gamma$ of $La_{0.87}Sm_{0.13}FeAs_{0.91}P_{0.09}O$, defined as the ratio of $H_{c2}^{//ab}$ to $H_{c2}^{//c}$, determined at a given temperature with the 50% $\rho_n$ criterion, increases from $\sim 4$ at 13.8 K to $\sim 7$ at 12 K. Such temperature dependent $\gamma$ further supports a multi-band superconductivity scenario, where the different Fermi-surface sheets develop distinct gaps in the superconducting state [58].

It is important to note the significant difference between the upper critical field values $H_{c2}$ and the respective anisotropy determined via magnetometry or resistivity measurements. The origin of this difference lies most likely in the inhomogeneous character of chemical pressure. Indeed, besides inducing an overall compression of the structure - leading to the appearance of superconductivity, it may selectively expand certain regions of the unit cell - leading to a variation of superconducting-state parameters. The upper critical field $H_{c2}$ determined via magnetization measurements corresponds rather to the bulk properties of the studied material, whereas $H_{c2}$ determined from resistivity measurements reflects the properties of its percolation network. Consequently, due to the inhomogeneous character of chemical pressure, the values of the superconducting parameters may differ significantly, depending on the employed technique. In this respect, the outcome of bulk techniques, such as magnetization or specific-heat measurements, is to be preferred over the less reproducible resistivity data.

Comparing the $H_{c2}$ anisotropy of $Ln$1111 crystals just below $T_c$ is not an easy task, especially when different experimental techniques are used (see above). Firstly, the anisotropy of $H_{c2}$ is particularly sensitive to the chosen criteria (see, for example, Fig. 15 in Ref. 59). Secondly, the $Ln$1111 [30] and La$_{0.87}$Sm$_{0.13}$FeAs$_{0.91}$P$_{0.09}$O crystals, especially those grown by the NaAs flux method, are thicker than crystals grown by the NaCl/KCl flux [23-26]; thus, increasing the possibility of defects in the *ab* planes. This can broaden the superconducting transition, as well as affect the curvature of $H_{c2}$(T) near $T_c$, which could mask possible two-band effects.

In summary, the superconducting-state parameters obtained from the magnetization measurements correspond to the average thermodynamic parameters of the studied materials. At the same time, the respective values obtained from resistivity measurements correspond to the upper limit of the thermodynamic parameters with a properly tuned and homogenous distribution of pressure across the full sample volume.

*3.6. Electronic band structure*

Electronic band-structure calculations are expected to provide important hints regarding the changes in the electronic properties of LaFeAsO upon simultaneous Sm and P substitution. Figure 9 shows the calculated orbital-resolved density of states (DOS) and the band structure of La$_{0.87}$Sm$_{0.13}$FeAs$_{0.91}$P$_{0.09}$O. The electronic structure of La$_{0.87}$Sm$_{0.13}$FeAs$_{0.91}$P$_{0.09}$O is characterized by two sets of disconnected Fermi-surface sheets, with two hole sheets of cylindrical shape, at the center of the Brillouin zone ($\Gamma$-point), and two electron pockets centered at $(0,\pm\pi)$ and $(\pm\pi,0)$ (*M*-point). Since the Sm and P co-doping does not bring additional electrons to the FeAs planes, the size and the shape of the electron- and hole pockets are substantially the same as in LaFeAsO, *i.e.*, they are two dimensional and mostly cylindrical, as illustrated in Fig. 9.

We recall that, the electronic structure of LaFeAsO and the influence of doping on its Fermi surface and magnetism were intensively studied by theoretical methods [60-62]. Contrary to earlier reports, where the calculated (*i.e.*, not measured) value of the internal atomic position $Z_{As}$ was used, the band structure of the fluorine-doped LaFeAsO$_{1-x}$F$_x$ is mostly two dimensional and almost insensitive to doping. Now, there is a consensus that DFT calculations, which are mean-field by nature, underestimate the effect of spin fluctuations, that generally suppress the long-range magnetic order [63,64]. Moreover, in our calculations the effect of disorder due to chemical substitution at random sites is not considered. Disorder is most likely a detrimental factor to the long-range magnetic order, favoring other magnetic phases characterized by the presence of spin fluctuations. A comprehensive understanding of the pairing mechanism in the La$_{0.87}$Sm$_{0.13}$FeAs$_{0.91}$P$_{0.09}$O system would most likely require a combination of experimental techniques and theoretical models that take into account all these factors. The predicted Fermi surface contains hole- and electron pockets of similar size (nesting), which favour an extended s-wave pairing, but do not rule out other possible pairing symmetries, including s-wave and d-wave.

Our recent NMR measurements of LaFe(As,P)O [65], as well as other studies [66,67], support the view that spin fluctuations play an important role in the superconducting pairing. Detailed calculations in the F-doped LaFeAsO materials showed their proximity to a quantum critical point, with an anomalously flat energy landscape, implying that even weak perturbations can induce significant changes in the physical properties [68]. In LaFeAsO, the most obvious way to induce such a quantum-critical transition is the isoelectronic substitution of As by P and of La by Sm. Indeed, the smaller ionic radii of P and Sm lead to a smaller cell volume and, hence, to an enhanced kinetic energy and to reduced electronic correlations. Thus, the onset of superconductivity in the La$_{0.87}$Sm$_{0.13}$FeAs$_{0.91}$P$_{0.09}$O compound can be readily interpreted in terms of the general picture widely accepted for the Fe-based superconductors [67].

## 4. Conclusions

To summarize, by using high-temperature high-pressure conditions, we could successfully synthesize a new kind of a mixed rare-earth iron-based oxide superconductor (La,Sm)Fe(As,P)O, where we substitute simultaneously La with Sm and As with P. The resulting single-crystalline samples, with partial substitutions in both the charge-reservoir- and the iron layers, were thoroughly investigated in order to establish their crystallographic- and their basic superconducting properties. We show that, despite the lack of a formal doping, the isovalent substitution of La$^{3+}$ by Sm$^{3+}$ and As$^{3-}$ by P$^{3-}$ results in a decrease of the unit cell volume and the appearance of bulk superconductivity, most likely due to the disorder-induced enhancement of spin fluctuations. Our findings indicate that even small differences in chemical pressure can profoundly affect the ground state of the original system. Our extension of previous single-site-substitution studies to double-site isovalent substitutions suggests a new route to induce superconductivity in iron-based materials. The emergence of superconductivity in the rare-earth iron-based oxides by means of simultaneous chemical pressure opens up new possibilities in the search for new classes of superconductors, where chemical pressure is essential to the onset of superconductivity.

**Acknowledgements**

We thank P. Macchi and S. Katrych for their help during the structural characterization and I. Mazin for helpful comments on the electronic band-structure calculations. T.S. was in part supported by the Swiss National Science Foundation under Grant No. 200021-169455.

**Figures and captions**

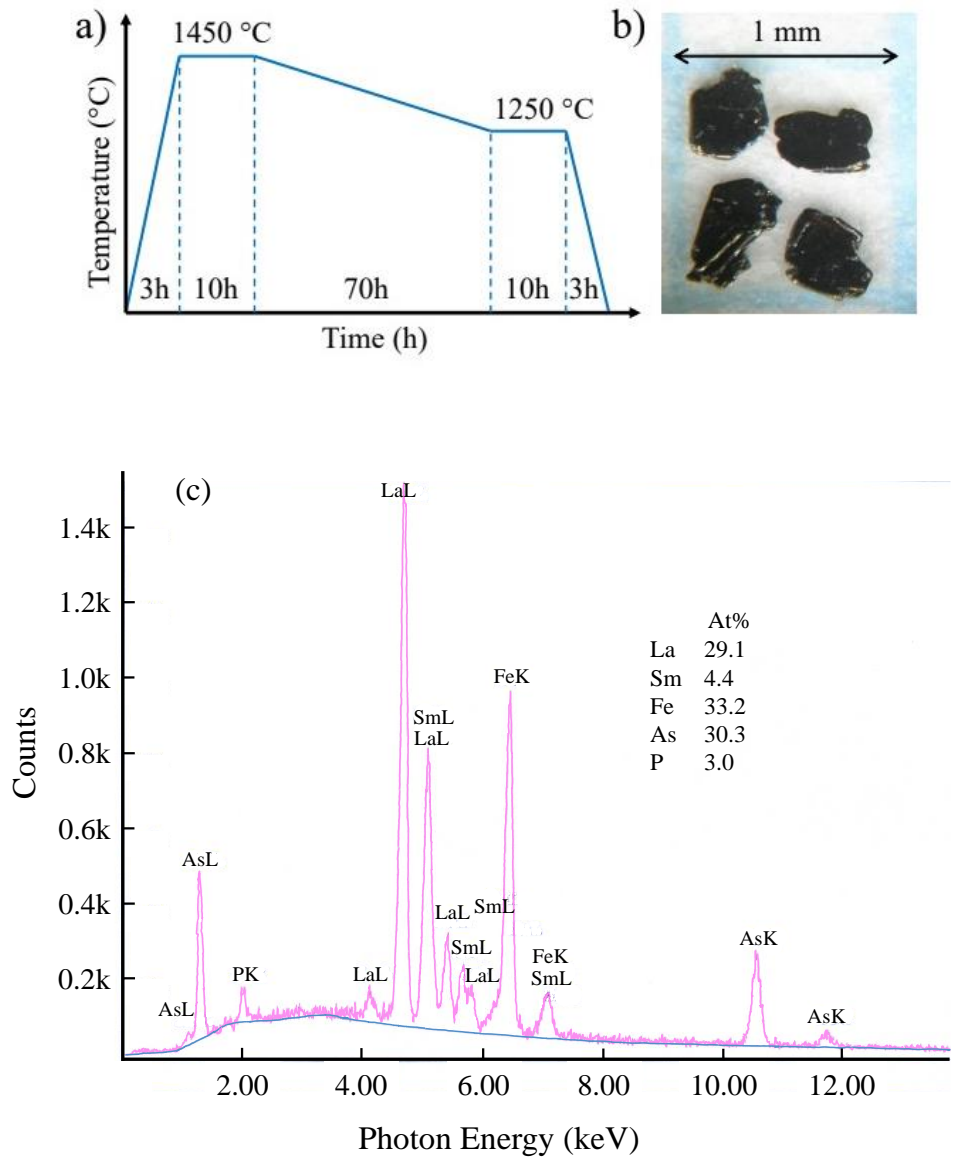

**Fig. 1.** (a) Heat-treatment protocol used for the high-pressure and high-temperature synthesis of (La,Sm)Fe(As,P)O crystals. (b) Optical microscope image of (La,Sm)Fe(As,P)O individual crystals. (c) EDX spectrum of a single $La_{0.87}Sm_{0.13}FeAs_{0.91}P_{0.09}O$ crystal.

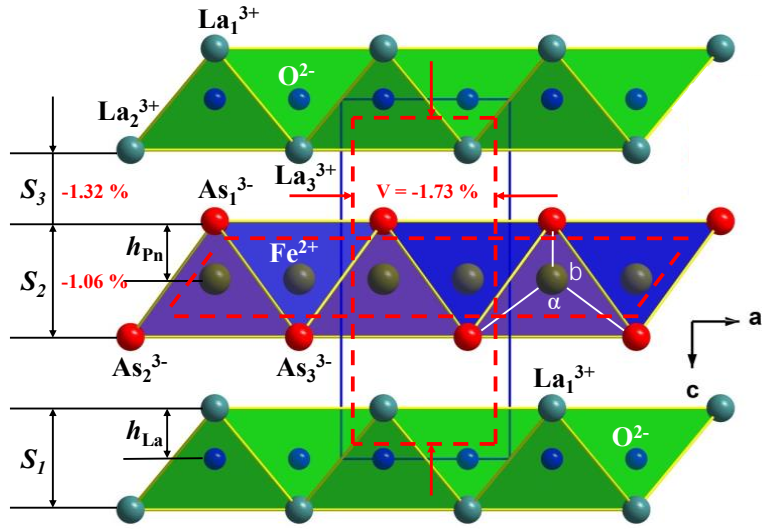

**Fig. 2.** Schematic representation of the LaFeAsO lattice projected on the *ac* plane (for details see Table 1). Red dotted lines and digits highlight the changes in lattice dimensions with the substitution of La by Sm and As by P. Blue and green polyhedra indicate the Fe(As,P)$_4$ and (La,Sm)O$_4$ tetrahedra, stacked alternately along the *c* axis. Here, $h_{Pn}$ is the height of the As/P atoms above the Fe plane, while $h_{Ln}$ is the height of the rare-earth metal atom above the O plane. $S_1$ is the thickness of the charge-supplying layer; $S_2$ is the thickness of the conducting layer, while $S_3$ is the interlayer distance. The As-Fe-As bond angles α and β (right-side tetrahedron) illustrate the deviation from a regular FeAs$_4$ tetrahedron, whose α and β angles are both 109.47°.

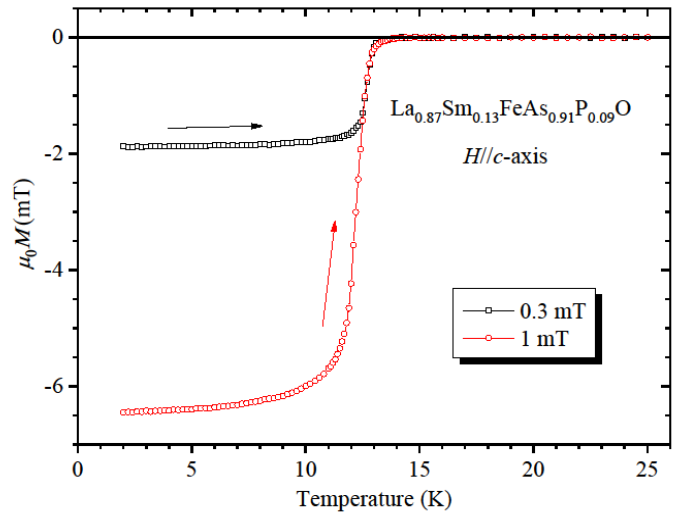

**Fig. 3.** Temperature-dependent dc magnetization measured in zero-field-cooled (ZFC) mode at 0.3 and 1 mT, with the field applied parallel to the *c*-axis of a La$_{0.87}$Sm$_{0.13}$FeAs$_{0.91}$P$_{0.09}$O crystal. The sharp superconducting transition at $T_c$ = 13.3 K indicates a good crystal quality, while the strong diamagnetic response well below $T_c$ is consistent with bulk superconductivity.

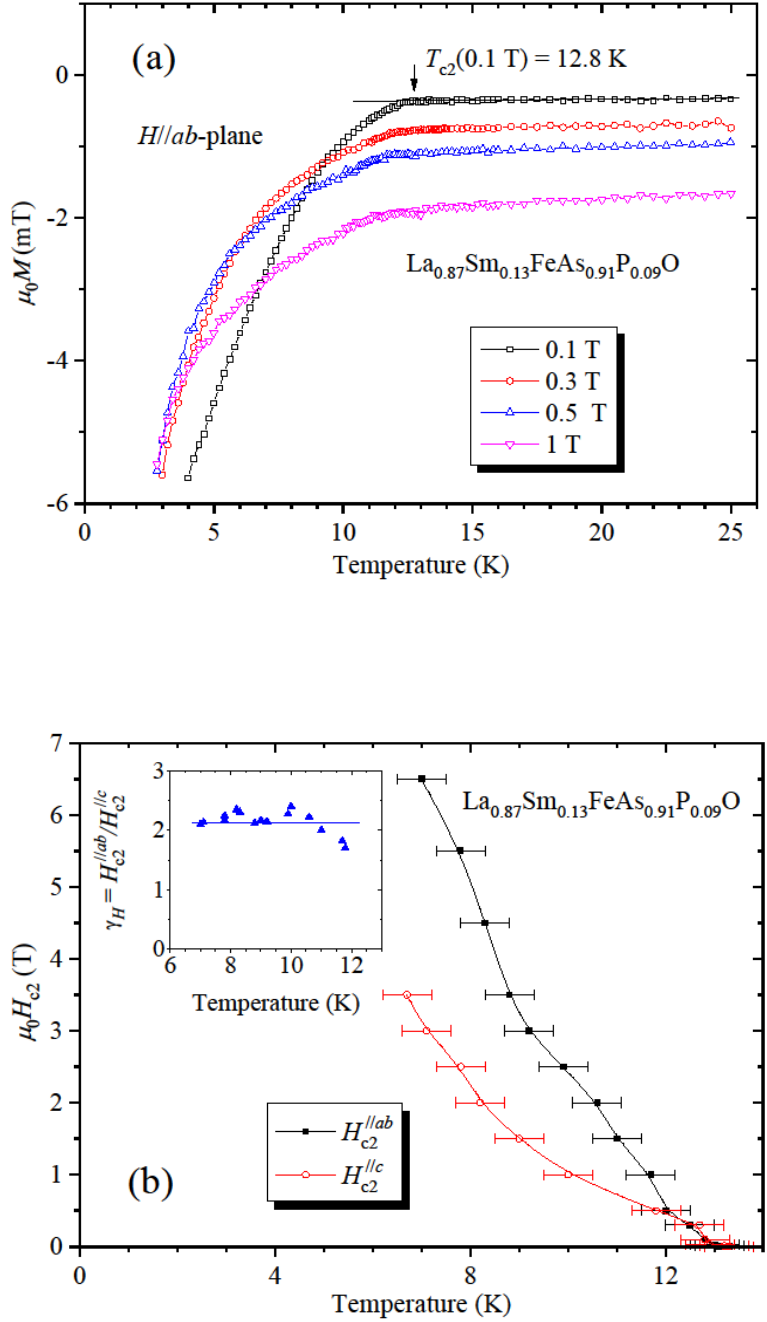

**Fig. 4.** (a) Temperature dependence of magnetization for selected $H//ab$-plane magnetic fields. (b) Temperature dependence of the upper critical field for $H//c$-axis and for $H//ab$-plane. The inset shows the temperature dependence of the anisotropy of the upper critical field, here defined as the ratio of $H_{c2}^{//ab}$ to $H_{c2}^{//c}$ at a given temperature.

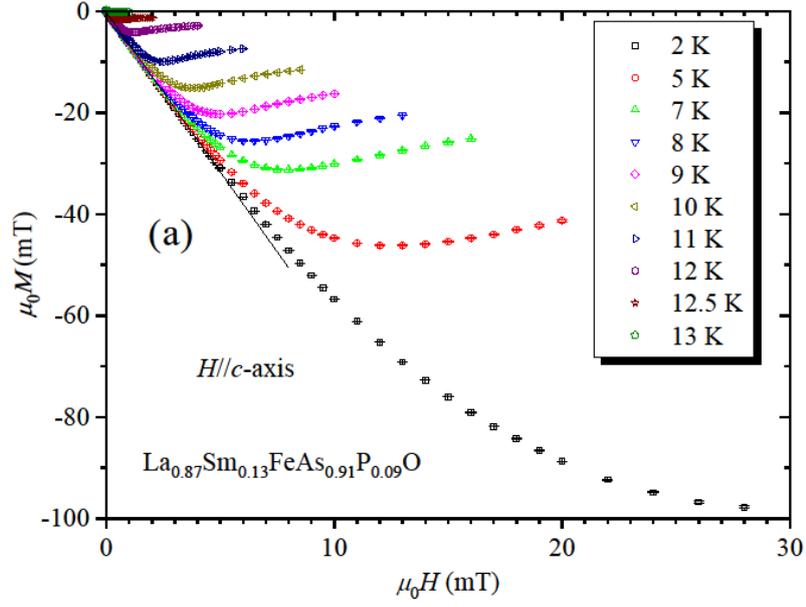

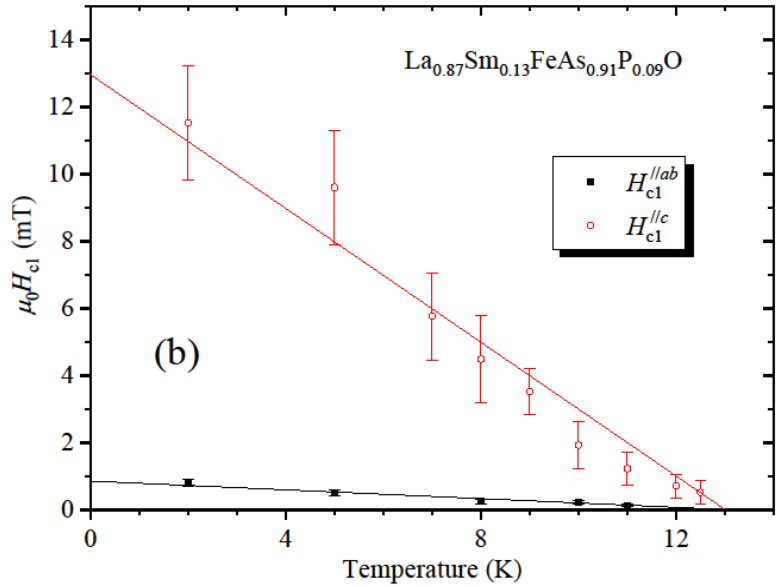

**Fig. 5.** (a) Initial magnetization vs magnetic field, for *H*//*c*-axis, recorded at selected temperatures. The zero-field value of the -d*M*/d*H* derivative is significantly larger than 1 due to the large demagnetizing field in the studied geometry. (b) Temperature dependence of the lower critical field, for *H*//*ab*-plane and *H*//*c*-axis, determined as described in the text.

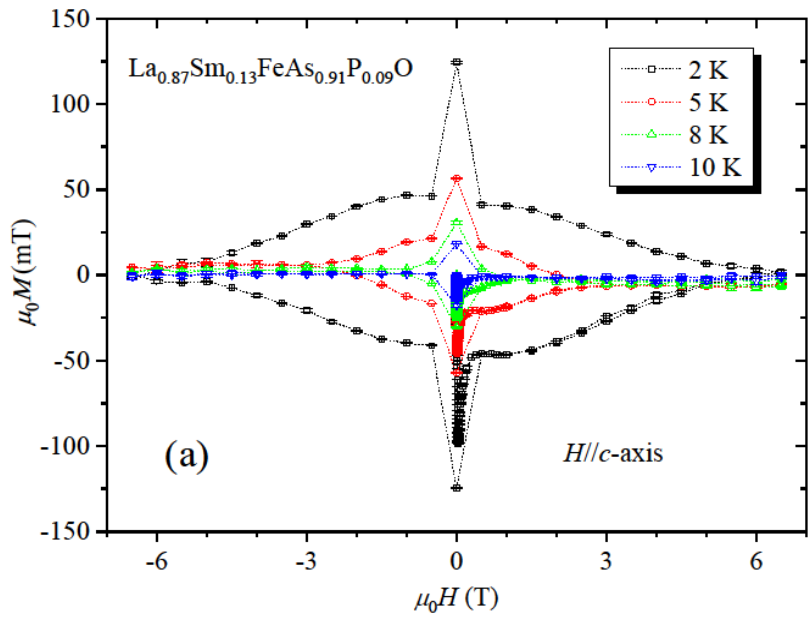

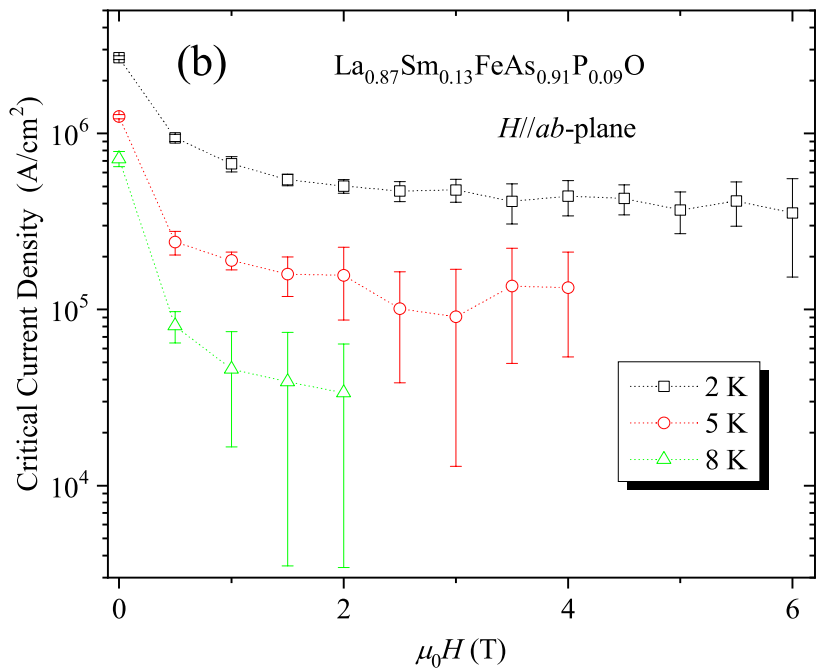

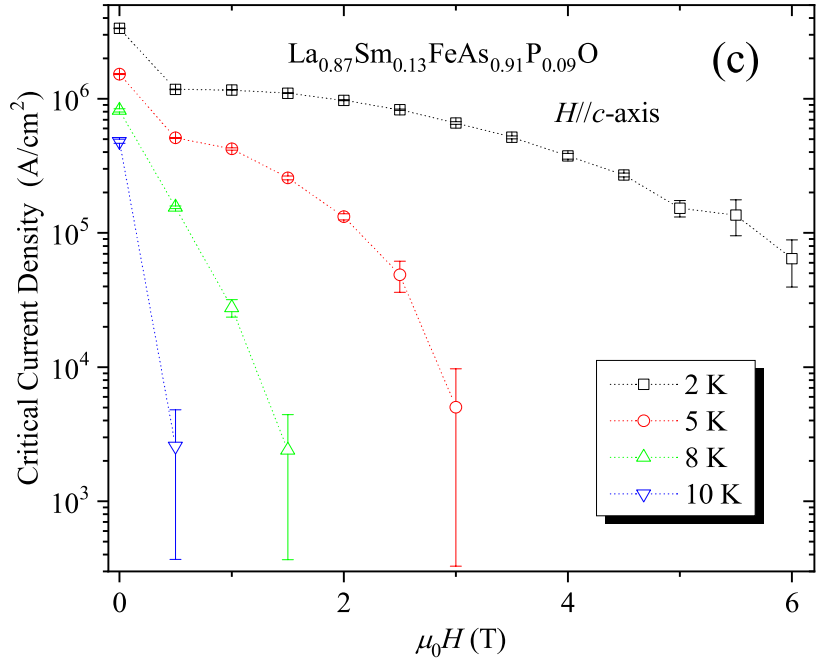

**Fig. 6.** Hysteresis loops for *H*//*c*, as recorded at selected temperatures (a). Critical-current density vs. magnetic field, as recorded at various temperatures, for *H*//*ab*-plane (b) and *H*//*c*-axis (c).

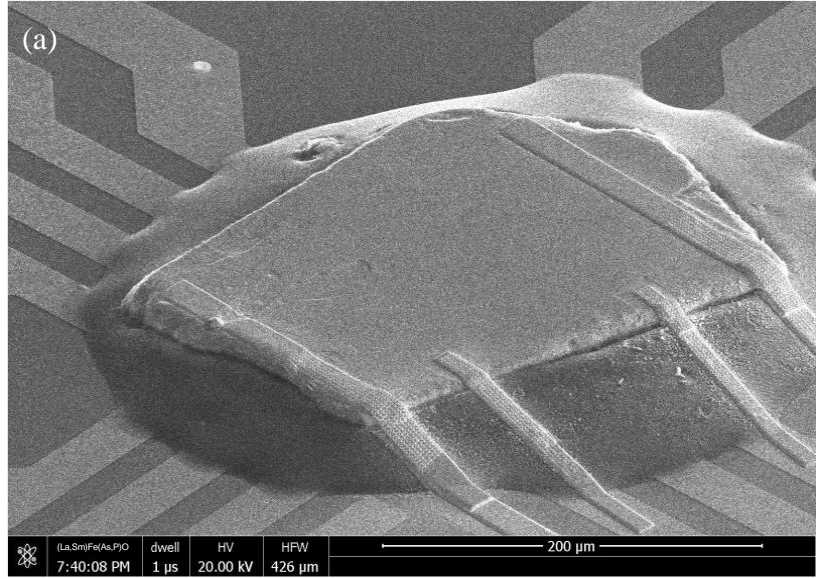

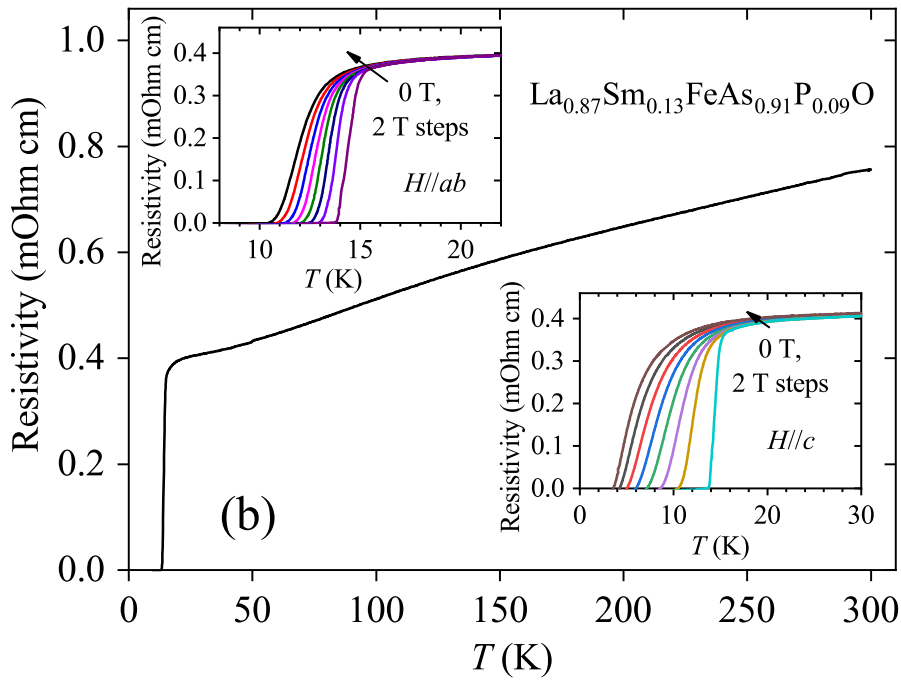

**Fig. 7.** (a) Scanning-electron-microscopy image of a La$_{0.87}$Sm$_{0.13}$FeAs$_{0.91}$P$_{0.09}$O single crystal showing the four platinum leads deposited by means of a focused-ion beam. (b) Temperature dependence of the resistance of a La$_{0.87}$Sm$_{0.13}$FeAs$_{0.91}$P$_{0.09}$O single crystal grown under high-pressure conditions. Left and right insets highlight the $T_c$ region for a series of magnetic fields (0, 2, 4, 6, 8, 10, 12, and 14 T) applied parallel to the Fe$_2$(As,P)$_2$ layers ($H$//$ab$ - top) and perpendicular to them ($H$//$c$ - bottom), respectively.

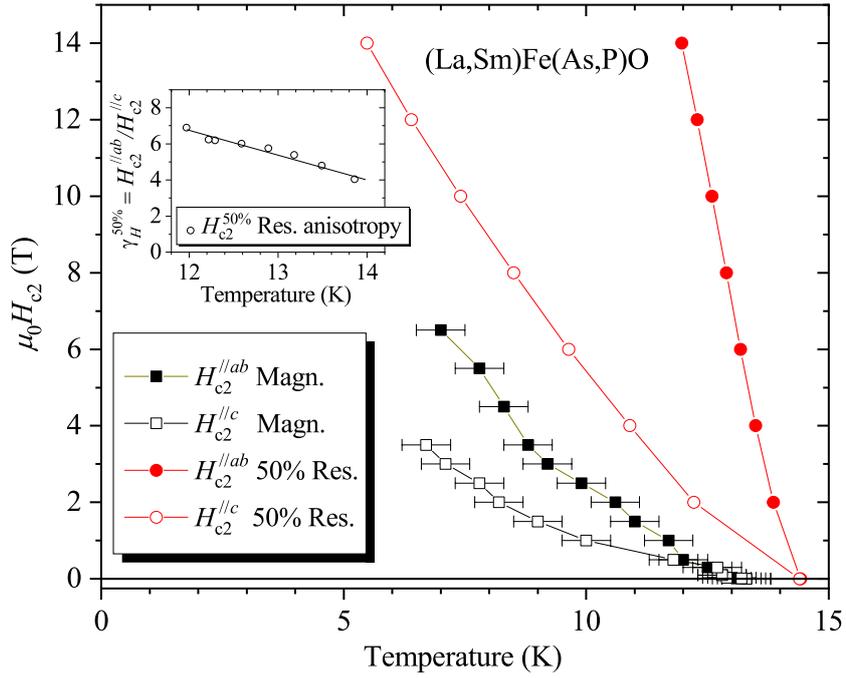

**Fig. 8.** Temperature dependence of the upper critical field for *H//ab* and *H//c* determined from magnetometry and resistivity measurements. To determine $H_{c2}$ from resistivity measurements, the 50% $\rho_n$ criterion was used. The inset shows the temperature dependence of the anisotropy of the upper critical field from resistivity measurements, defined as the ratio of $H_{c2}^{//ab}$ to $H_{c2}^{//c}$ at a given temperature.

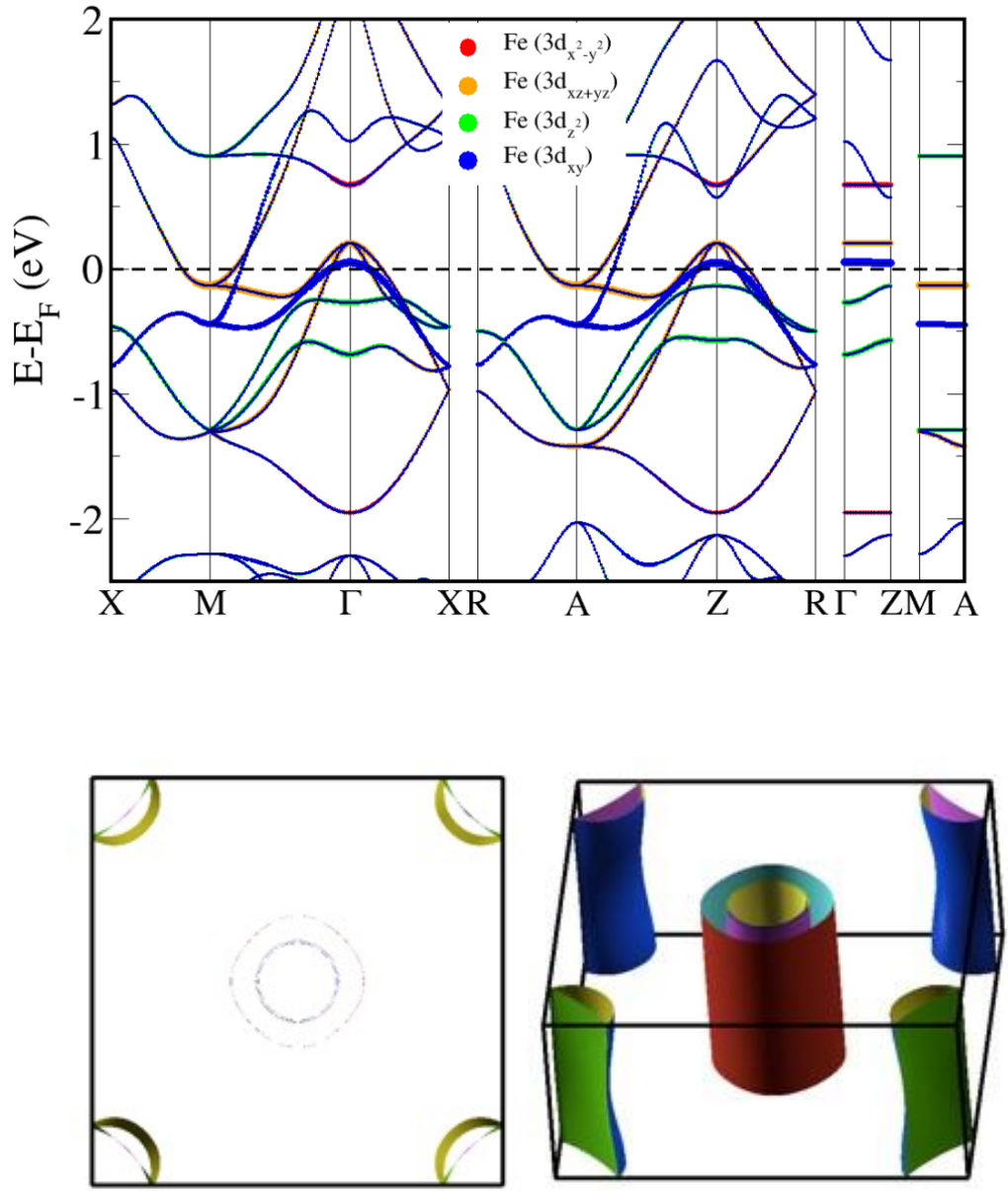

**Fig. 9.** Electronic band structure near the Fermi energy and the Fermi-surface configuration for $La_{0.87}Sm_{0.13}FeAs_{0.91}P_{0.09}O$ (top and perspective views), as resulting from DFT calculations. Colors highlight the orbital-resolved Fe-3$d$ character of the states.

**Table 1.** Crystallographic and structural refinement parameters for a $La_{0.87}Sm_{0.13}FeAs_{0.91}P_{0.09}O$ single crystal ($T$ = 295 K, Mo $K_\alpha$, $\lambda$ = 0.71073 Å). The absorption effects were corrected analytically. A full-matrix least-squares method was used to optimize the goodness-of-fit parameter $F^2$. The lattice is tetragonal, with a space group $P4/nmm$ (No. 123). The uncertainties in the last digits are given in parentheses. Some distances and marking of atoms are shown in Fig. 2. The structural parameters for LaFeAsO are taken from Refs. 16 and 39.

| Chemical formula | $La_{0.87}Sm_{0.13}FeAs_{0.91}P_{0.09}O$ | LaFeAsO |
|---|---|---|
| $T_c$ (K) | 13.3 | non SC |
| Unit cell dimensions (Å) | $a$ = 4.0179(2) | $a$ = 4.0357(3) |
|  | $c$ = 8.6630(7) | $c$ = 8.7378(6) |
| Cell volume (Å$^3$) | 139.851(15) | 142.31(2) |
| $z_{La/Sm}$ (atomic coordinate) | 0.1424(1) | 0.1411(2) |
| $z_{As/P}$ | 0.6510(1) | 0.6513(3) |
| La/Sm$_1$-La/Sm$_2$ (Å) | 3.7625(4) | |
| O-O = Fe-Fe (Å) | 2.8411(1) | 2.8528(1) |
| La/Sm$_2$-As/P$_2$ (Å) | 3.3581(6) | 3.3836(12) |
| La/Sm-O (Å) | 2.3573(3) | 2.3682(16) |
| As/P$_1$-As/P$_2$ (Å) | 3.8618(9) | |
| Fe-As/P (Å) | 2.3972(5) | 2.4049(20) |
| As/P$_1$-Fe-As/P$_2$, $\beta$ (deg.) | 107.318(19) | |
| As/P$_2$-Fe-As/P$_3$, $\alpha$ (deg.) | 113.87(4) | 113.5(1) |
| $S_3$ (Å) | 1.790(1) | 1.8139 |
| $S_1$ (Å) | 2.467(2) | 2.466(2) |
| $h_{Pn}$ ($S_2/2$) (Å) | 1.3081(9) | 1.322 |
| Calculated density (g/cm$^3$) | 6.575 | |
| Absorption coefficient (mm$^{-1}$) | 29.432 | |
| $F(000)$ | 241 | |
| Crystal size (μm$^3$) | 209 × 130 × 22 | |
| $\theta$ range for data collection | 2.35° - 51.11° | |
| Index ranges | $-8 \leq h \leq 8$, $-8 \leq k \leq 4$, $-16 \leq l \leq 18$ | |
| Reflections collected/unique | 2617/484 $R_{int.}$ = 0.0456 | |
| Completeness to $2\theta$ | 96.0 % | |
| Data/restraints/parameters | 484/0/12 | |
| Goodness of fit on $F^2$ | 1.177 | |
| Final $R$ indices [$I > 2\sigma(I)$] | $R_1$ = 0.0429, w$R_2$ = 0.1015 | |
| $R$ indices (all data) | $R_1$ = 0.0465, w$R_2$ = 0.1029 | |
| $\Delta\rho_{max}$ and $\Delta\rho_{min}$, (e/Å$^3$) | 15.496 and -3.037 | |

**Table 2.** Atomic coordinates and the equivalent isotropic- and anisotropic displacement parameters (Å$^2$ × 10$^3$) for La$_{0.87}$Sm$_{0.13}$FeAs$_{0.91}$P$_{0.09}$O. Here, $U_{iso}$ is defined as one third of the trace of the orthogonalized $U_{ij}$ tensor. In the anisotropic displacement factor, the exponent takes the form: $-2\pi^2(h^2a^2U_{11} + \ldots + 2hka^*b^*U_{12})$. For symmetry reasons $U_{23} = U_{13} = U_{12} = 0$.

| Atom | Site | x | y | z | $U_{iso}$ | $U_{11} = U_{22}$ | $U_{33}$ |
|---|---|---|---|---|---|---|---|
| La/Sm | 2c | -¼ | -¼ | 0.1424(1) | 9(1) | 6(1) | 14(1) |
| Fe | 2b | ¼ | ¾ | ½ | 11(1) | 8(1) | 16(1) |
| As/P | 2c | ¼ | ¼ | 0.3491(2) | 11(1) | 9(1) | 14(1) |
| O | 2a | ¼ | ¾ | 0 | 10(1) | 8(2) | 13(2) |

**Table 3.** Thermodynamic parameters describing the superconducting state of a La$_{0.87}$Sm$_{0.13}$FeAs$_{0.91}$P$_{0.09}$O single crystal.

| Parameter | Value |
|---|---|
| $T_c$ (K) | 13.3(2) |
| $-d\mu_0 H_{c2}^{//c}/dT$ (T K$^{-1}$) | 0.76(8) |
| $-d\mu_0 H_{c2}^{//ab}/dT$ (T K$^{-1}$) | 1.2(2) |
| $\mu_0 H_{c2}^{//c}(0)$ (T) | 7(1) |
| $\mu_0 H_{c2}^{//ab}(0)$ (T) | 11(2) |
| $\mu_0 H_{c1}^{//c}(0)$ (mT) | 13(2) |
| $\mu_0 H_{c1}^{//ab}(0)$ (mT) | 0.85(9) |
| $\xi_{ab}(0)$ (nm) | 6.9(9) |
| $\xi_c(0)$ (nm) | 4.4(6) |
| $\lambda_{ab}(0)$ (nm) | 225(30) |
| $\lambda_c(0)$ (nm) | 5000(700) |
| $\kappa^{//c}(0) = \lambda_{ab}(0)/\xi_{ab}(0)$ | 33(4) |
| $\kappa^{//ab}(0) = [\lambda_{ab}(0)\lambda_c(0)/\xi_{ab}(0)\xi_c(0)]^{1/2}$ | 195(40) |